\documentclass[11pt]{article}

\usepackage[final]{acl}
\author{
  Jun Gao$^1$ \quad Yun Peng$^3$ \quad Qian Qiao $^7$\quad  Changhai Zhou$^5$  \quad Yuhua Zhou$^1$\\ \textbf{Shiyang Zhang$^6$} \quad \textbf{Shichao Weng}$^5$ \quad \textbf{Zhenchang Xing$^4$} \quad \textbf{Xiaoxue Ren$^{12}$}\thanks{*Corresponding Author}\\
  $^1$ School of Software Technology, Zhejiang University \\
  $^2$ Hangzhou High-Tech Zone (Binjiang) Institute of Blockchain and Data Security \\
  $^3$ Chinese University of Hong Kong 
  $^4$ CSIRO's Data61\\
  $^5$ Fudan University 
  $^6$ Yale University 
  $^7$ Independent Researcher \\
  \texttt{\{jgao1106, xxren\}@zju.edu.cn}
}

\usepackage{times}
\usepackage{latexsym}

\usepackage[T1]{fontenc}

\usepackage[utf8]{inputenc}

\usepackage{microtype}

\usepackage{inconsolata}

\usepackage{graphicx}
\usepackage{multirow}
\usepackage{booktabs}
\usepackage{subcaption}
\usepackage{adjustbox}
\usepackage{bbding}
\usepackage{colortbl}
\usepackage{booktabs} 
\usepackage{xcolor}   
\usepackage{array}    
\usepackage{amsmath}
\usepackage{amssymb}
\usepackage{url}
\usepackage{tcolorbox}
\usepackage{listings}

\usepackage[utf8]{inputenc}
\usepackage{tcolorbox}
\usepackage{listings}
\usepackage{xcolor}
\usepackage{textcomp} 

\tcbuselibrary{breakable, skins}

\lstdefinelanguage{json}{
    basicstyle=\small\ttfamily,
    columns=fullflexible,
    breaklines=true,
    stringstyle=\color{blue},
    literate=
     *{0}{{{\color{orange}0}}}{1} {1}{{{\color{orange}1}}}{1} {2}{{{\color{orange}2}}}{1}
      {3}{{{\color{orange}3}}}{1} {4}{{{\color{orange}4}}}{1} {5}{{{\color{orange}5}}}{1}
      {6}{{{\color{orange}6}}}{1} {7}{{{\color{orange}7}}}{1} {8}{{{\color{orange}8}}}{1}
      {9}{{{\color{orange}9}}}{1},
}

\newtcolorbox{promptbox}[1]{
    breakable,
    enhanced,
    colback=gray!3!white,
    colframe=gray!75!black,
    fonttitle=\bfseries,
    title=#1,
    arc=0mm,
    left=3mm,
    right=3mm,
    top=3mm,
    bottom=3mm,
    before skip=12pt,
    after skip=12pt
}
\definecolor{highlightgray}{rgb}{0.95, 0.95, 0.95}

\title{CoRE: A Fine-Grained Code Reasoning Benchmark Beyond Output Prediction}
\begin{document}
\maketitle
\begin{abstract}
Despite strong performance on code generation tasks, it remains unclear whether large language models (LLMs) genuinely reason about code execution. 
Existing code reasoning benchmarks primarily evaluate final output correctness under a single canonical implementation, leaving two critical aspects underexplored:
(1) whether LLMs can maintain consistency to functionally equivalent implementations, and (2) whether LLMs can accurately reason about intermediate execution states.
We introduce \textbf{CoRE}, a \textbf{Co}de \textbf{Re}asoning benchmark that evaluates code reasoning through \textbf{implementation invariance} and \textbf{process transparency}.
Extensive evaluations on eight frontier LLMs reveal two fundamental limitations. 
First, models exhibit a substantial \textbf{robustness gap}, with performance varying significantly across equivalent implementations. 
Second, we observe \textbf{superficial execution}, where models arrive at correct final outputs without correctly reasoning about intermediate execution states.
Together, these findings demonstrate that output-only evaluations are insufficient for assessing code reasoning and position CoRE as a necessary benchmark for evaluating robust and faithful code reasoning.\footnote{Data and code are available at https://github.com/ZJUSig/CoRE.}
\end{abstract}

\section{Introduction}
\begin{figure}[!t]
    \centering
    \includegraphics[width=0.98\linewidth]{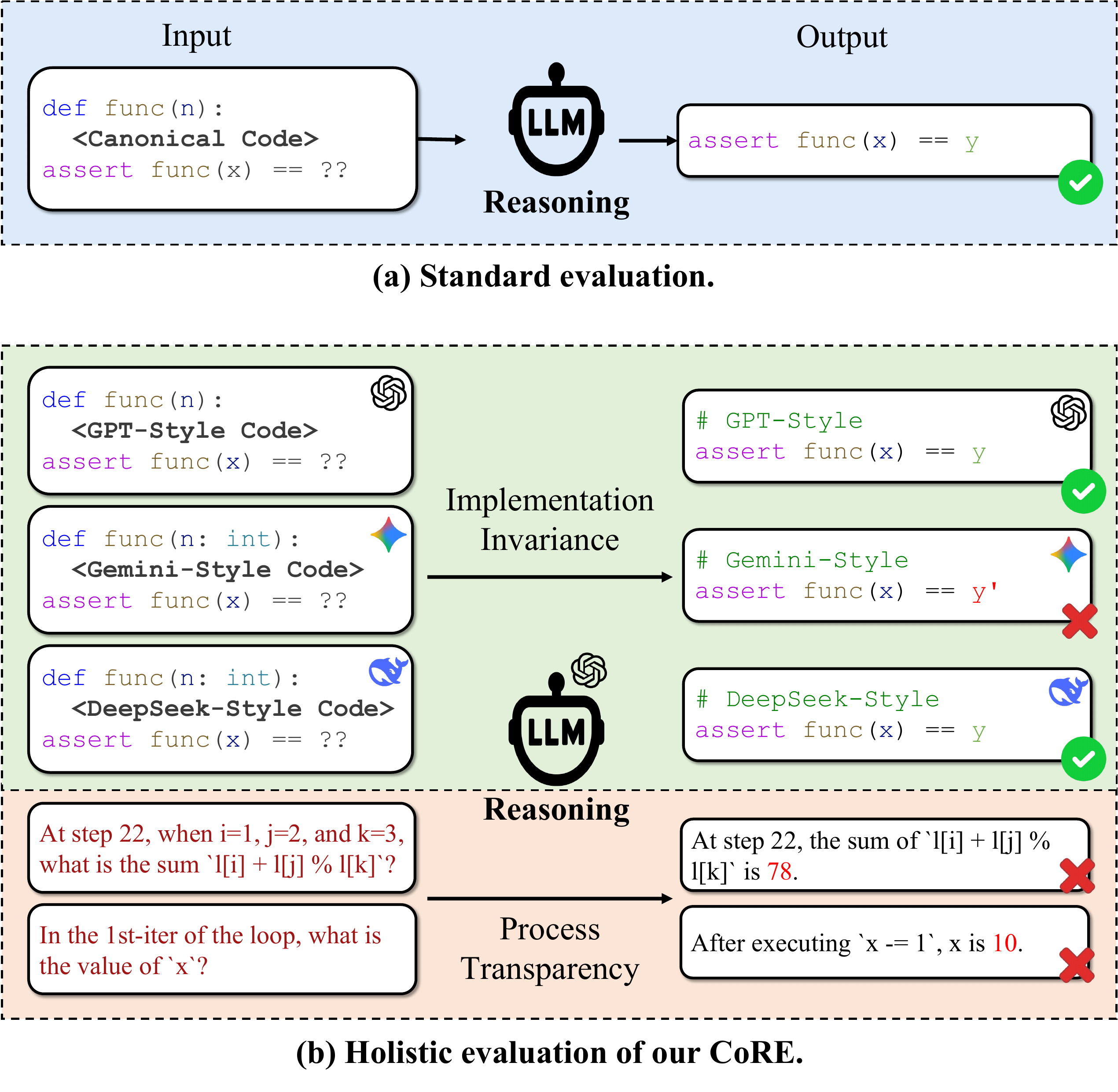}
    \caption{The code reasoning comparison of standard code reasoning evaluation and our holistic evaluation with diverse implementations and intermediate probing. 
}
    \label{fig:difference}
\end{figure}
\begin{figure*}[!t]
    \centering
    \begin{minipage}[c]{0.32\textwidth}
        \centering
        \includegraphics[width=\linewidth]{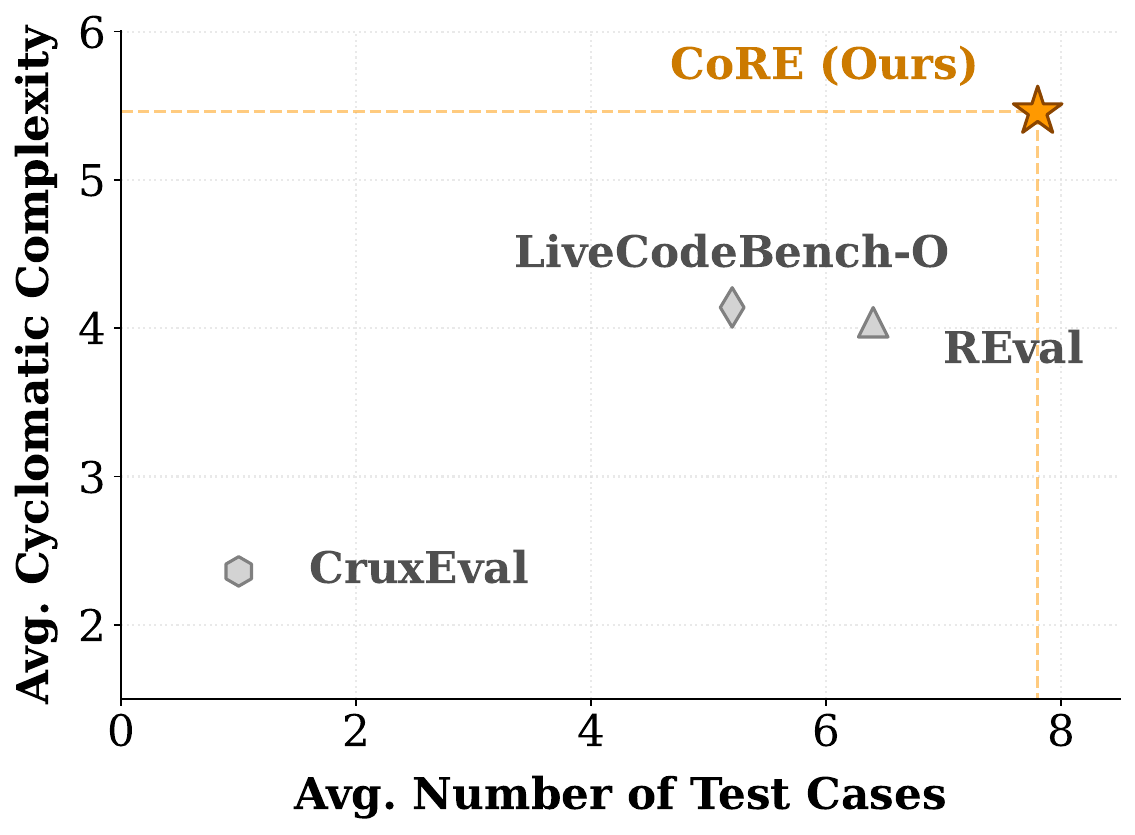}
    \end{minipage}
    \hfill 
    \begin{minipage}[c]{0.67\textwidth}
        \centering
        \small 
        \setlength{\tabcolsep}{3.5pt} 
        
        \begin{tabular}{lcccccc}
            \toprule    
            &  \textbf{Samples}  & \textbf{Tests}  & \textbf{Impl.} & \textbf{ $\mathcal{J}^\downarrow$} & \textbf{CC $^\uparrow$} & \textbf{Inter. $^\uparrow$} \\
            \midrule
            CruxEval & 800  & 1.0 & 1.0 & 1.0 & 2.4 & \XSolidBrush\\
            LiveCodeBench-O & 478  & 5.2 & 1.0 & 1.0 & 4.1 & \XSolidBrush\\
            REval  & 955  & 6.4 & 1.0 & 1.0 & 4.0 & 3.0 \\
            \midrule
            CoRE (ours) & \textbf{1,978}  & \textbf{7.8}& \textbf{4.3} & \textbf{0.6} & \textbf{5.0} & \textbf{4.1} \\
            \bottomrule
        \end{tabular}
    \end{minipage}

\caption{
Comparison of \textbf{CoRE} against existing code reasoning benchmarks.
(Left) Distribution of Average Cyclomatic Complexity and Average Number of Test Cases.
(Right) Comparison of code reasoning benchmarks.
\textbf{Samples}: Total number of samples for inference.
\textbf{Tests}: Average number of test cases per instance.
\textbf{Impl.}: Average number of diverse code implementations per coding problem.
\textbf{$\mathcal{J}$}: Average Jaccard similarity between diverse implementations for each coding problem, where lower indicates higher diversity.
\textbf{CC}: Average Cyclomatic Complexity of code implementations.
\textbf{Inter.}: Average number of intermediate state probes per coding problem.
}
\label{fig:highlight}
\end{figure*}
The capabilities of Large Language Models (LLMs) in code-relevant tasks have evolved rapidly from simple code completion to solving complex programming problems~\cite{hurst2024gpt, liu2024deepseek, yang2024swe, bouzenia2024repairagent, sun2025don, guo2025deepseek,zhong2025anytalker, zhou2026balancingfidelityplasticityaligning}.
However, recent studies indicate that the ability to generate syntactically correct code does not necessarily imply a genuine understanding of its execution~\cite{zhaounveiling,gao2025texttt}.
This discrepancy raises a fundamental question: \textit{Do LLMs truly understand the execution logic or are they merely relying on superficial heuristics?} 

In this paper, we argue that current benchmarks are insufficient due to two critical flaws.
First, they lack \textbf{implementation invariance}, defined as the ability to evaluate reasoning robustness across functionally equivalent but structurally or lexically distinct code implementations.
However, benchmarks like CruxEval~\cite{gu2024cruxeval}, LiveCodeBench-O~\cite{jainlivecodebench}, and REval~\cite{chen2025reasoning} primarily rely on a single \textit{canonical} solutions, as illustrated in Fig.\ref{fig:difference}(a).
Second, existing benchmarks lack \textbf{process transparency}, as they typically predict the final outputs without verifying intermediate execution states.
While REval pioneered the exploration of intermediate reasoning states, it remains limited by a rigid templating paradigm and a fundamental disregard for implementation invariance.

\begin{figure*}[!t]
    \centering
    \begin{minipage}[c]{0.69\textwidth}
        \includegraphics[width=0.98\linewidth]{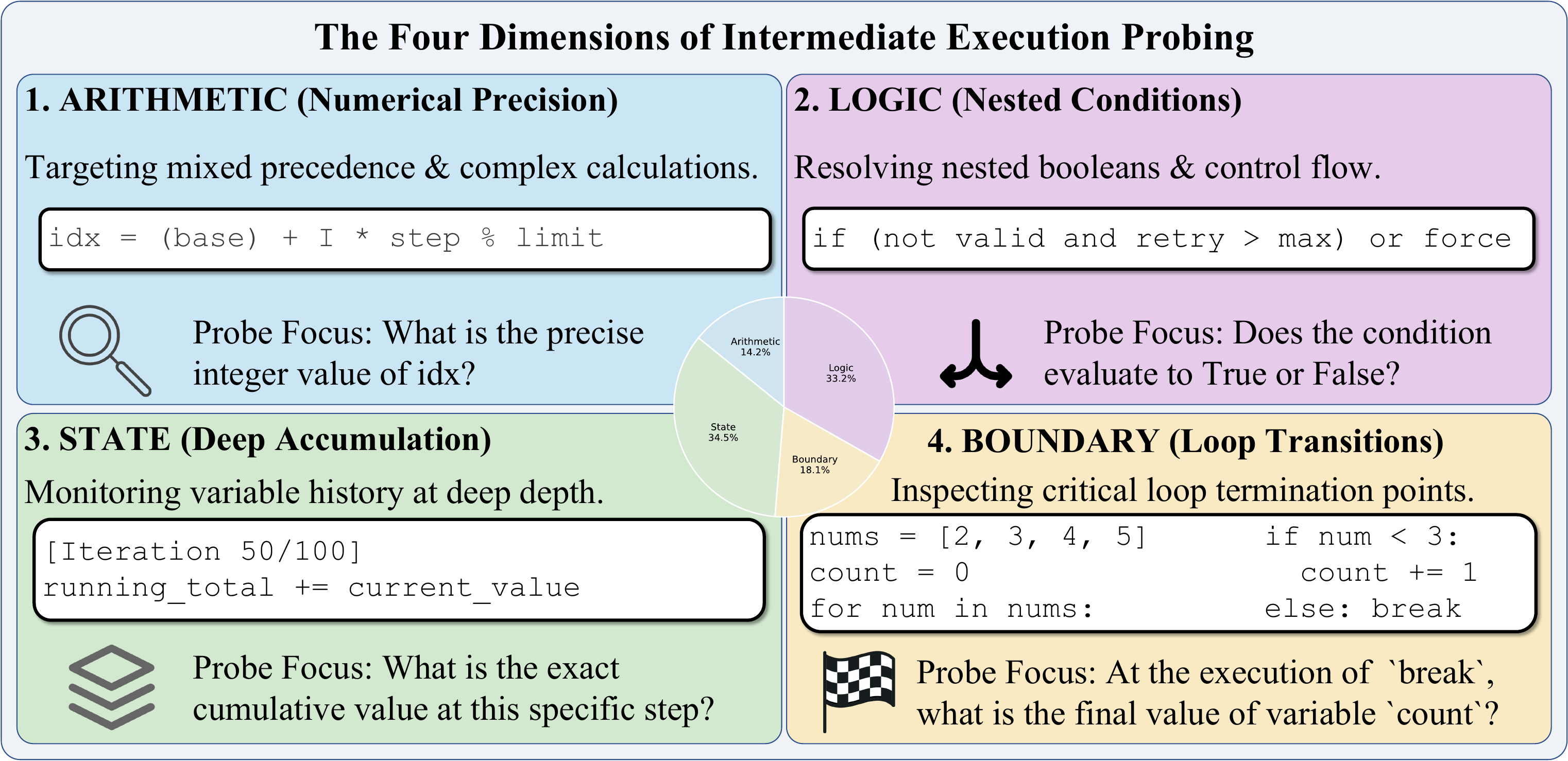}
    \end{minipage}
    \hfill 
    \begin{minipage}[c]{0.3\textwidth}
        \centering
        \small 
        \renewcommand{\arraystretch}{1.20}

        \resizebox{\textwidth}{!}{
        \begin{tabular}{llc }
            \toprule    
            & \textbf{Aspect} & \textbf{Number}   \\
            \midrule
            \multirow{4}{1.2cm}{Basic Statistic} & Code Problems & 60 \\
            & Code Candidates & 255 \\
            & Intermediate State Probes & 243 \\
            & Samples & 1978 \\
            \midrule
            \multirow{5}{1.2cm}{Impl. Statistic} & OpenAI &  74\\
            & DeepSeek & 68\\
            & Gemini & 51\\
            & Claude & 41 \\
            & Qwen & 21 \\
            \midrule
            \multirow{5}{1.2cm}{Probe Statistic} &  Arithmetic & 77\\ 
            & Logic & 178\\
            & State & 185 \\
            & Boundary & 97 \\
            & Avg. Dimensions & 2.2\\
            \bottomrule

        \end{tabular}}
    \end{minipage}

    \caption{The taxonomy and distribution of Intermediate Probing dimensions.
    The left panel illustrates the four dimensions of execution probing.
    The right table summarizes the dataset composition.
    The middle section details the distribution of code candidates generated by various LLMs, and the bottom specifies the number of intermediate probing questions assigned to each reasoning dimension.
    }
    \label{fig:statistic}
\end{figure*}

In light of these limitations, we introduce CoRE, a \textbf{Co}de \textbf{Re}asoning benchmark that jointly evaluates \textbf{implementation invariance} and \textbf{process transparency} in code reasoning.
CoRE is constructed from 60 coding problems from HumanEval~\cite{chen2021evaluating} and LiveCodeBench~\cite{jain2024livecodebench}, selected for their diverse test cases and algorithmic complexity.
For each problem, CoRE evaluates whether LLMs produce consistent predictions across diverse but functionally equivalent implementations.
As illustrated in Fig.\ref{fig:difference}(b), we leverage implementations generated by the various LLMs listed in Tab.\ref{tab:model_selection} to quantify implementation invariance.
Beyond this, CoRE further evaluates process transparency by examining the model's capacity to reason about intermediate execution states.
To this end, we utilize verified codebases and employ five LLMs listed in Tab.\ref{tab:model_selection} to generate probes targeting intermediate states within nested loops and complex conditional branches.
These intermediate state probes cover four empirical reasoning dimensions, \textit{Arithmetic, Logic, State, and Boundary}, and are designed to test whether LLMs follow faithful step-by-step reasoning rather than relying on superficial heuristics.
In total, as shown in Fig.\ref{fig:highlight} and Fig.\ref{fig:statistic}, CoRE comprises 255 unique code implementations, with an average of 4.1 implementations per problem for implementation invariance evaluation.
Each coding problem contains 7.8 test cases on average, and each implementation exhibits an average cyclomatic complexity of 5.0, as shown in Fig.\ref{fig:highlight}.
For process transparency, each problem includes an average of 4.1 intermediate state probes, each covering 2.2 reasoning dimensions on average.

Our evaluation of eight frontier LLMs, including GPT-5, o3, Claude-4.5, and DeepSeek-V3.2, reveals critical limitations in code reasoning.
First, we identify a significant \textbf{Robustness Gap}, where performance is inconsistent across syntactically diverse but functionally identical codes.
Specifically, we observe that LLMs typically exhibit the highest proficiency with code generated by their own model family, and second-best on code from the OpenAI series, indicating an overfitting where LLMs are biased toward familiar styles.
Second, we observe \textbf{Superficial Execution}, where LLMs produce correct function outputs but hallucinate intermediate states, revealing reliance on superficial heuristics rather than genuine execution understanding.

Overall, our contributions are threefold: (1) We propose CoRE, a challenging benchmark designed to rigorously assess code reasoning through implementation invariance and process transparency. (2) We reveal that LLMs frequently fail due to stylistic overfitting and superficial execution. (3) We introduce a holistic evaluation protocol that exposes the fragility of current LLMs, demonstrating that genuine execution understanding lags significantly behind output prediction.
\begin{figure*}[!t]
    \centering
    \includegraphics[width=0.98\linewidth]{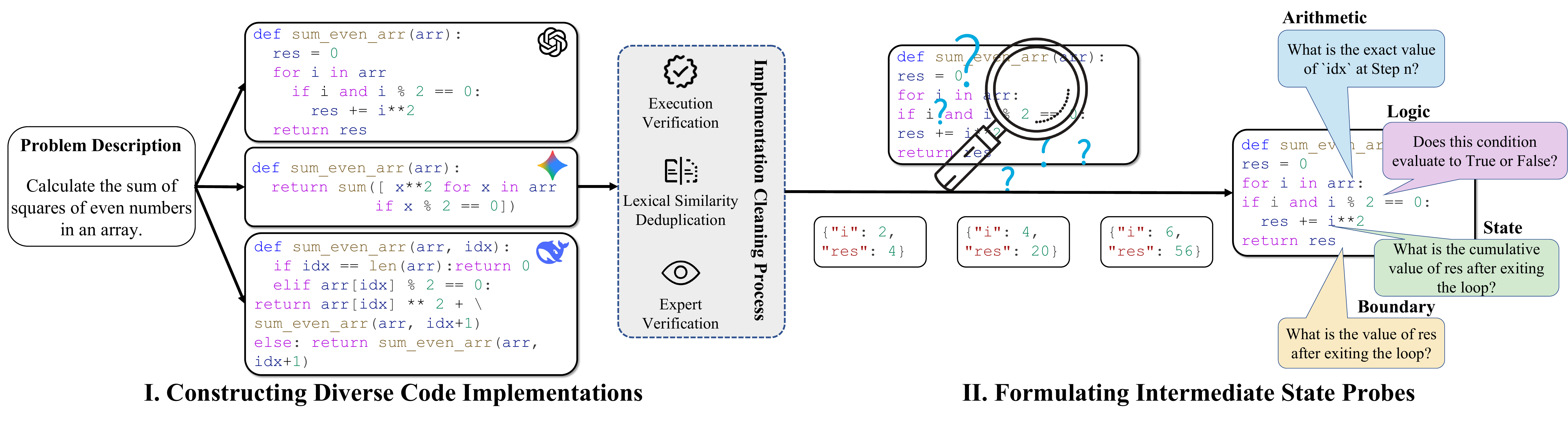}
    \caption{The construction pipeline of CoRE.
    It consists of two stages: (I) \textbf{Constructing Diverse Code Implementations}, which produces functionally equivalent code via massive LLMs, and validates its functionality and diversity, and (II) \textbf{Formulating Intermediate State Probes}, where LLMs are employed to synthesize probes grounded in captured execution traces across four dimensions: \textit{Arithmetic, Logic, State, and Boundary}.}
    \label{fig:method_workflow}
\end{figure*}
\section{Related Studies}

Code reasoning tasks have emerged as a rigorous method to assess execution understanding in LLMs~\cite{liu2024codemind,gao2025texttt,zhaounveiling}.
Although benchmarks like CruxEval~\cite{gu2024cruxeval} and LiveCodeBench-O~\cite{jain2024livecodebench} establish a foundation for this domain, they prioritize output prediction and effectively treat the reasoning mechanism as a black box.
The reliance on single canonical implementations and low-complexity code makes it difficult to distinguish genuine execution simulation from heuristic input-output mapping.
This distinction is vital given that LLM reasoning remains fragile and often unfaithful.
Even with Chain-of-Thought prompting~\cite{wei2022chain}, models frequently generate plausible yet hallucinated traces derived from shallow heuristics rather than actual logic reasoning~\cite{liu2024codemind, beger2025coconut, gao2025aim, lanham2023measuring, turpin2023language, wang2024large, gao2025interleaved,ji2025specvlm, zhou2024qprunerprobabilisticdecisionquantization,zhou2026lara,zhou2025bslora}.
Such unreliability necessitates validating intermediate states instead of relying solely on terminal outputs~\cite{uesato2022solving, lightman2023let}.
While REval~\cite{chen2025reasoning} attempts to incorporate intermediate state prediction, its dependence on rigid templates and simple canonical code restricts its ability to evaluate reasoning robustness across varied coding styles.
Current methods fail to verify if LLMs maintain consistency when processing functionally equivalent but syntactically diverse solutions.
CoRE addresses these limitations by enforcing implementation invariance and process transparency to provide a strict evaluation of code reasoning.

\section{CoRE Benchmark}
In this section, we detail the construction process of CoRE, which aims to evaluate both the implementation invariance and process transparency of LLMs in code reasoning.
The construction pipeline is shown in Fig.\ref{fig:method_workflow}, which consists of two stages.
The first stage constructs diverse code implementations, and the second stage formulates intermediate state probes.

\subsection{Constructing Diverse Code Implementations}
\label{sec:div}
To establish a high-quality foundation for code reasoning, we first collect an initial coding problem set from HumanEval~\cite{chen2021evaluating} and LiveCodeBench~\cite{jain2024livecodebench}. 
These benchmarks are selected for their rigorous coverage of test cases and algorithmic complexity, which can be used for evaluating code reasoning.
We initially aggregate a pool of problems $\mathcal{P}$ comprising 164 coding problems from HumanEval and 880 from LiveCodeBench.
As illustrated in first stage of our construction workflow in Fig.\ref{fig:method_workflow}, we then leverage seven diverse LLMs $\mathcal{M}$ to generate potential solutions, formulating a comprehensive pool $\mathcal{S}$ containing 7,308 distinct code implementations for 1,044 unique problems, defined as:
\begin{equation}
    \mathcal{S} = \{ s \mid s \sim m(p), \forall p \in \mathcal{P}, \forall m \in \mathcal{M} \}.
\end{equation}

To ensure that the benchmark effectively evaluates code reasoning, we apply a hierarchical implementation cleaning process.
We first perform execution validation to ensure functional equivalence, retaining only the solutions that pass their corresponding test sets $\mathcal{T}_{\mathcal{P}}$.
We further refine this set to guarantee reasoning depth and lexical variety.
Specifically, an instance $s$ is retained if and only if its cyclomatic complexity $\mathcal{C}(s)$ exceeds a threshold $\tau_{\text{cc}}$ and its 1-gram Jaccard similarity $\mathcal{J}$ with other implementations \textit{sharing the same problem ID} remains below $\tau_{\text{sim}}$.
The 1-gram Jaccard similarity~\cite{JaccardIndex} are formulated as:
\begin{equation}
    \mathcal{J}(s, t) = \frac{| \mathcal{G}_1(s) \cap \mathcal{G}_1(t) |}{| \mathcal{G}_1(s) \cup \mathcal{G}_1(t) |},
\end{equation}
where $\mathcal{G}_1(s)$ denotes the set of unique 1-grams (tokens) in code implementation $s$.
Overall, the implementation cleaning process is defined as:
\begin{equation}
\mathcal{S}' = \left\{ s \;\middle|\;
\begin{aligned}
    & \mathcal{C}(s) \ge \tau_{\text{cc}} \\
    & \land \max_{\substack{t \in \mathcal{D}_{\text{io}} \setminus \{s\} \\ \text{id}(s) = \text{id}(t)}} \mathcal{J}(s, t) \le \tau_{\text{sim}}
\end{aligned}
\right\},
\end{equation}

Recognizing that evaluating LLMs on such diverse candidates with massive test suites presents significant computational challenges, we optimize evaluation efficiency by computing $\mathcal{T_P}^*$, the minimal subset of test cases achieving maximal coverage for problem $\mathcal{P}$.
This reduction strategy reduces the average number of test cases per instance from 24.1 to 7.8 without compromising coverage integrity.
Therefore, the larger number of test cases reported in Fig.\ref{fig:highlight} is not a superficial increase in quantity, but reflects test cases that are necessary to ensure adequate coverage.
Finally, human expert verification ensures implementation diversity, producing 255 curated implementations across 60 problems, as shown in Fig.\ref{fig:highlight}.

\subsection{Formulating Intermediate State Probes}
\label{sec:probes}
To formulate intermediate state probes, we first employ Python execution tracing to capture the runtime values of all variables, as illustrated in the middle of Fig.\ref{fig:method_workflow}, serving as ground truth for process transparency evaluation.
Then, we employ an ensemble of LLMs to automatically produce intermediate state probes targeting four critical dimensions of program behavior: \textit{Arithmetic, Logic, State, and Boundary}.
Specifically, the \emph{Arithmetic} dimension tracks numerical precision in composite operations and complex indexing, e.g., \texttt{res = (a + b) * c \% d}.
The \emph{Logic} dimension evaluates the resolution of compound 
boolean conditions and nested control flow decisions.
To ensure reasoning depth, the \emph{State} dimension monitors variable histories across long execution paths, whereas the \emph{Boundary} dimension focuses on critical transition points, such as loop terminations.
Notably, to increase challenge and reasoning depth, we ask LLMs to prioritize probes that span multiple dimensions, for example, \textit{tracking state accumulation within nested logical branches}.
Finally, human experts validate the intermediate probes for logical correctness, diversity, and complexity to ensure the rigor of CoRE.
This process yields an intermediate state probe set for each problem $\mathcal{P}$, denoted as $\mathcal{Q_P}$, which contains an average of 4.1 probes, with each probe spanning 2.2 dimensions on average.

\subsection{Benchmark Statistics Analysis}
\textbf{Basic statistics.}
CoRE expands existing benchmarks by an order of magnitude with 1,978 samples.
It raises the evaluation complexity, with an average cyclomatic complexity of 5.0 and an average of 7.8 test cases per coding problem, markedly surpassing existing benchmarks, as shown in Fig.\ref{fig:highlight}.

\noindent\textbf{Implementation Invariance.}
As detailed in the basic statistic in Fig.\ref{fig:statistic}, CoRE consists of 60 coding problems, which contain a total of 255 code candidates, namely, an average of 4.3 implementations per coding problem.
The diversity of the dataset is demonstrated in the Impl. Statistics in Fig.\ref{fig:statistic}, comprising 74 from OpenAI, 51 from Gemini, 68 from DeepSeek, 41 from Claude, and 21 from Qwen.
The inner diversity of implementations within each coding problem is evidenced by a low 1-gram Jaccard similarity score of $\mathcal{J}=0.6$ as listed in Fig.\ref{fig:highlight}.

\noindent\textbf{Process Transparency.}
To evaluate LLMs in Process Transparency, we also formulate intermediate probes for each coding problem across four different dimensions.
As shown in probe statistic in Fig.\ref{fig:statistic}, 77 probes involving \textit{Arithmetic}, 178 involving \textit{Logic}, 185 involving \textit{State}, and 97 involving \textit{Boundary}.
These dimensions are not mutually exclusive, and they are often combined.
For example, the \textit{State} dimension is frequently interacted with others as illustrated in Fig.\ref{fig:upset}.
On average, each individual probe covers 2.2 dimensions, ensuring that the evaluation effectively challenges LLMs' process reasoning.

\section{Evaluation Protocol}
\label{sec:eval}
Unlike traditional evaluations, which typically focus on the output accuracy of a single canonical implementation, our protocol provides a holistic assessment of code reasoning by leveraging the diverse implementations $\mathcal{S_P}$ and intermediate probes $\mathcal{Q_P}$ constructed in Sec.\ref{sec:div} and Sec.\ref{sec:probes}.

\subsection{Preliminaries}
We first recall some background in code reasoning evaluation in this section.

\noindent\textbf{Standard Code Reasoning.} Traditional benchmarks~\cite{chen2021evaluating,jain2024livecodebench} primarily focus on output prediction using a canonical Implementation $c$.
This approach treats the reasoning process as a black box, where the model is evaluated on its ability to map an input $x$ to the execution result $\hat{y}$ for the given $c$:
\begin{equation}
\hat{y} = \text{Pred}(c, x),
\end{equation}
where $\hat{y}$ is expect to match the ground truth $y$.

\noindent\textbf{Intermediate State Reasoning.}
Benchmarks such as REval~\cite{chen2025reasoning} frame intermediate-state reasoning as a question-answering task, covering path reachability and variable prediction.
Formally, given a code implementation $c$, a intermediate probes $p$, and a specific test input $x$, the model predicts the intermediate state as: 
\begin{equation} 
\hat{a} = \text{Pred}(p, c, x),
\end{equation}
where $\hat{a}$ is the predicted answer to the intermediate probes $p$.

\subsection{Implementation Invariance}
A robust model should maintain consistency across diverse yet functionally equivalent implementations.
For a given problem $\mathcal{P}$, we define its diverse code implementations in CoRE as a set $\mathcal{S_{P}}=\{c_{1}, c_{2}, ..., c_{n}\}$.
Implementation invariance is quantified via the \textbf{Strict Output Consistency $I$}, a binary indicator that is satisfied only if the model $\mathcal{M}$ correctly predicts the final output for every candidate solution across all associated test cases $\mathcal{T}_{\mathcal{P}}$:
\begin{equation}
I(\mathcal{P}) = 
\begin{cases} 
1, & \forall c_i \in \mathcal{S}_{\mathcal{P}},  \mathcal{M}(c_i, \mathcal{T_P}^x) =\mathcal{T_P}^y \\
0, & \text{otherwise}.
\end{cases}
\end{equation}
Additionally, we also define the soft accuracy $I_s$ across multiple code implementations within each question $\mathcal{P}$:
\begin{equation}
    I_s(\mathcal{P}) = \frac{\sum_{c_i \in \mathcal{S_P}} \mathcal{M} (c_i, \mathcal{T_{P}}^x) == \mathcal{T_P}^y}{|\mathcal{S_P}|} 
\end{equation}

\subsection{Process Transparency}
Moreover, reliable LLMs are expected to effectively demonstrate process transparency in code reasoning rather than simple input-output mapping.
We define $\mathcal{Q}_{\mathcal{P}} = \{(q_1, a_1), (q_2, a_2), \dots, (q_m, a_m)\}$ as the probing question set of question $\mathcal{P}$.
Process transparency of code reasoning is calculated by \textbf{Process Fidelity Weight} $W_s$, the ratio of probing questions correctly answered by the model $\mathcal{M}$:
\begin{equation}
W_{s}(\mathcal{P}) = \frac{\sum_{(q_i, a_i) \in \mathcal{Q}_{\mathcal{P}}} \mathbb{1}(\mathcal{M}(q_i, \mathcal{P}) = a_i)}
{|\mathcal{Q}_{\mathcal{P}}|},
\end{equation}
where $\mathbb{1}(\cdot)$ is the indicator function, and $|\mathcal{Q}_{\mathcal{P}}|$ represents the total number of probes for the problem $\mathcal{P}$.
Additionally, we define a strict fidelity indicator $W \in \{0, 1\}$, where $W = 1$ if and only if all intermediate probing questions in $\mathcal{Q_P}$ are answered correctly, i.e., $W_{s} = 1$, and $W = 0$ otherwise.

\subsection{Reasoning Consistency Score}
To rigorously quantify the robustness of code reasoning and penalize superficial heuristics, we propose the \textbf{Reasoning Consistency Score (RCS)}.
The final RCS for problem $\mathcal{P}$ is computed as the product of the strict output consistency and the process fidelity weight:
\begin{equation}
RCS(\mathcal{P}) = I \times W_{s}
\end{equation} 
This formulation ensures that \textit{any} failure in implementation invariance nullifies the final score, while scaling that score by the depth of its internal execution probing.
By coupling these two dimensions, the RCS effectively filters out shortcut reasoning, distinguishing genuine code reasoning from a reliance on superficial heuristics.

\begin{table*}[!t]
  \centering
  \setlength{\aboverulesep}{0pt}
  \setlength{\belowrulesep}{0pt}
  \resizebox{\textwidth}{!}{ 
  \begin{tabular}{l cccccc cccccc}
  \toprule
  \multirow{2}{*}{\textbf{Method}} & \multicolumn{2}{c}{Impl. Invar.} & \multicolumn{2}{c}{Proc. Trans.} & \multicolumn{2}{c}{Overall} & \multicolumn{2}{c}{Impl. Invar.} & \multicolumn{2}{c}{Proc. Trans.} & \multicolumn{2}{c}{Overall} \\
  \cmidrule(lr){2-3} \cmidrule(lr){4-5} \cmidrule(lr){6-7} \cmidrule(lr){8-9} \cmidrule(lr){10-11} \cmidrule(lr){12-13}
  & \textbf{$I$} & \textbf{$I_{s}$} & \textbf{$W$} & \textbf{$W_{s}$} & \textbf{$RCS$} & \textbf{$Cons.$} & \textbf{$I$} & \textbf{$I_{s}$} & \textbf{$W$} & \textbf{$W_{s}$} & \textbf{$RCS$} & \textbf{$Cons.$} \\
  \midrule
  \rowcolor{highlightgray} \multicolumn{13}{l}{\textbf{OpenAI Series}} \\
  \rowcolor{highlightgray} & \multicolumn{6}{c}{\textit{GPT-5}} & \multicolumn{6}{c}{\textit{o3}} \\
  IO      & 84.62 & 95.44 & 0.00 & 18.96 & 15.93 & 0.15 & 83.52 & 95.74 & 0.00 & 9.73 & 8.90 & 0.16 \\
  CoT     & \textbf{87.91} & \textbf{96.77} & \underline{26.37} & \underline{64.93} & \underline{55.75} & 0.27 & 82.42 & 94.74 & 28.57 & 67.60 & 54.25 & 0.31 \\
  CoC     & \underline{86.81} & \underline{96.04} & 20.88 & 59.14 & 50.24 & \underline{0.30} & \textbf{87.91} & \textbf{97.05} & \textbf{31.87} & 67.55 & 58.24 & \underline{0.35} \\
  RHDA-1  & \underline{86.81} & 95.84 & 19.78 & 59.49 & 51.70 & 0.29 & 82.42 & 94.45 & \textbf{31.87} & \textbf{70.46} & \underline{58.32} & \textbf{0.36} \\
  RHDA-2  & \underline{86.81} & 95.83 & \textbf{30.77} & \textbf{68.92} & \textbf{58.48} & \textbf{0.33} & \underline{85.71} & \underline{95.75} & \underline{29.67} & \underline{68.28} & \textbf{59.16} & 0.33 \\
  \midrule
  \rowcolor{highlightgray} \multicolumn{13}{l}{\textbf{DeepSeek Series}} \\
  \rowcolor{highlightgray} & \multicolumn{6}{c}{\textit{DeepSeek-V3.2}} & \multicolumn{6}{c}{\textit{DeepSeek-R1}} \\
  IO      & 56.04 & 88.04 & \underline{29.67} & \textbf{67.95} & 40.49 & \textbf{0.56} & 32.97 & 84.43 & \underline{24.18} & 62.93 & 20.60 & \textbf{0.58} \\
  CoT     & \underline{64.84} & \underline{90.38} & 25.27 & 64.76 & \underline{41.94} & 0.41 & 34.07 & 86.19 & 20.88 & 63.26 & 18.52 & 0.52 \\
  CoC     & 60.44 & 88.13 & 26.37 & 66.47 & 40.40 & \underline{0.48} & 50.55 & 90.01 & 21.98 & 62.60 & 32.36 & \underline{0.54} \\
  RHDA-1  & 62.64 & 88.50 & 27.47 & 66.67 & 41.48 & 0.45 & \underline{70.33} & \underline{90.85} & \textbf{25.27} & \textbf{65.29} & \underline{44.34} & 0.37 \\
  RHDA-2  & \textbf{71.43} & \textbf{91.18} & \textbf{30.77} & \underline{67.07} & \textbf{48.70} & 0.42 & \textbf{72.53} & \textbf{91.51} & \underline{24.18} & \underline{64.49} & \textbf{47.16} & 0.38 \\
  \midrule
  \rowcolor{highlightgray} \multicolumn{13}{l}{\textbf{Claude Series}} \\
  \rowcolor{highlightgray} & \multicolumn{6}{c}{\textit{Claude-4.5}} & \multicolumn{6}{c}{\textit{Claude-3.7}} \\
  IO      & 65.93 & \underline{89.75} & 25.27 & 62.40 & 39.45 & 0.37 & 12.09 & 67.17 & 20.88 & 61.90 & 6.81 & \textbf{0.69} \\
  CoT     & \underline{68.13} & 88.50 & 27.47 & 64.23 & 43.90 & 0.37 & \underline{56.04} & \underline{87.38} & 18.68 & 63.00 & 37.53 & 0.49 \\
  CoC     & 61.54 & 88.05 & 23.08 & 64.78 & 39.78 & \textbf{0.44} & 48.35 & 86.27 & \textbf{27.47} & \underline{64.76} & 31.48 & \underline{0.53} \\
  RHDA-1  & 65.93 & 89.39 & \underline{28.57} & \underline{65.71} & \underline{44.34} & \underline{0.43} & \underline{56.04} & 86.04 & \underline{24.18} & \textbf{65.15} & \underline{39.73} & \underline{0.53} \\
  RHDA-2  & \textbf{81.32} & \textbf{93.71} & \textbf{29.67} & \textbf{68.74} & \textbf{54.45} & 0.29 & \textbf{64.84} & \textbf{88.75} & 21.98 & 64.32 & \textbf{42.77} & 0.44 \\
  \midrule
  \rowcolor{highlightgray} \multicolumn{13}{l}{\textbf{Gemini \& Qwen Series}} \\
  \rowcolor{highlightgray} & \multicolumn{6}{c}{\textit{Gemini-2.5}} & \multicolumn{6}{c}{\textit{Qwen-3}} \\
  IO      & 41.76 & 86.83 & 1.10 & 27.73 & 12.11 & \underline{0.57} & 5.49 & 53.02 & \textbf{0.00} & 9.38 & 0.27 & \textbf{0.95} \\
  CoT     & 51.65 & \underline{88.04} & \textbf{30.77} & \textbf{64.74} & 33.13 & 0.51 & 12.09 & \underline{78.07} & \textbf{0.00} & 18.90 & 2.69 & 0.88 \\
  CoC     & 38.46 & 87.15 & 18.68 & 59.58 & 23.41 & \textbf{0.58} & 6.59 & 71.75 & \textbf{0.00} & 17.53 & 1.26 & \underline{0.93} \\
  RHDA-1  & \underline{61.54} & 87.82 & 21.98 & 62.88 & \underline{38.90} & 0.47 & \underline{16.48} & 76.39 & \textbf{0.00} & \textbf{23.99} & \underline{4.18} & 0.84 \\
  RHDA-2  & \textbf{71.43} & \textbf{91.78} & \underline{29.67} & \underline{64.54} & \textbf{44.25} & 0.38 & \textbf{47.25} & \textbf{84.22} & \textbf{0.00} & \underline{22.20} & \textbf{10.66} & 0.53 \\
  \bottomrule
  \end{tabular}
  
  }
\caption{Code Reasoning performance of baselines on the CoRE benchmark.
The best performance is \textbf{bolded} and the second best is \underline{underline}.
\textbf{Across all frontier LLMs, models exhibit preferences over different code implementations, and frequently produce correct outputs despite incorrect intermediate execution states.}}
\label{tab:main_results}

\end{table*}

\section{Experiments}

\subsection{Experimental Setup}
\noindent\textbf{LLMs.} 
We evaluate the code reasoning capabilities of eight leading LLMs, including GPT-5~\cite{openai2025gpt5}, o3~\cite{openai2025o3mini}, Claude-4.5~\cite{anthropic2025claude45}, DeepSeek-V3.2~\cite{liu2025deepseek}, DeepSeek-R1~\cite{guo2025deepseek}, and Qwen-3~\cite{yang2025qwen3}.

\noindent\textbf{Prompting Methods.} 
To provide a comprehensive assessment, we employ diverse prompting strategies including standard Input-Output (IO), Chain-of-Thought (CoT)~\cite{wei2022chain}, and Chain-of-Code (CoC)~\cite{lichain}.
We also evaluate the RHDA framework~\cite{zhaounveiling} as a strong reasoning baseline.
In this setting, RHDA-1 and RHDA-2 correspond to the framework configured with one and two iterations, respectively.
Details of the prompts and postprocessing are provided in Appendix~\ref{app:baseline}.

\noindent\textbf{Metrics.}
We report the metrics defined in Sec.~\ref{sec:eval}, including Strict Output Consistency ($I$), soft accuracy across implementations ($I_s$), Process Fidelity Weight ($W_s$), and strict process consistency ($W$).
Our primary metric is the Reasoning Consistency Score (RCS), which assesses whether correct outputs are supported by consistent intermediate-execution reasoning across functionally equivalent implementations.
We additionally report Cons., defined as $1-\text{MSE}(I, W_s)$, indicating how closely final output performance matches intermediate probe performance.
All experiments are performed in three times, and the results are presented as the average.

\subsection{Results}
Tab.\ref{tab:main_results} presents a comprehensive comparison of code reasoning performance across various LLM series and prompting methods.
We observe a significant performance gap, characterized by a disconnect between the reasoning step and the final prediction.
For instance, in the IO setting, GPT-5 achieves a high strict output accuracy $I$ of 84.62 but a low process fidelity weight $W_s$ of 18.96.
This discrepancy indicates that models often hallucinate intermediate states despite generating correct final answers, a phenomenon we term superficial execution.
Incorporating CoT substantially mitigates this issue, increasing $W_s$ to 64.93 for GPT-5, which demonstrates that explicitly reasoning improves process fidelity.
Regarding the reflection-based RHDA framework, implementation invariance is consistently enhanced, whereas gains in process transparency remain limited.
RHDA-2 even underperforms RHDA-1 in some cases.
This behavior reflects the inherent fragility of LLM reasoning.
When feedback is derived solely from outputs, surface-level discrepancies in results can influence previous reasoning steps, leading models to erroneously revise previously correct intermediate states.

Additionally, a performance gap between $I$ and $I_s$ implicitly suggests that models struggle to maintain consistency across different implementations, revealing a potential robustness gap.
A detailed analysis of this phenomenon is provided in Sec.~\ref{sec:robustness_gap}.
\begin{table*}[t] 
\centering 
\small 
\renewcommand{\arraystretch}{1.25}
\setlength{\tabcolsep}{4.5pt} 
\begin{tabular}{l @{\hskip 12pt} cccc @{\hskip 12pt} cccc @{\hskip 12pt} cccc} 
\toprule 
\multirow{2}{*}{\textbf{Model}} & \multicolumn{4}{c}{\textbf{IO}} & \multicolumn{4}{c}{\textbf{CoT}} & \multicolumn{4}{c}{\textbf{CoC}} \\ 
\cmidrule(lr){2-5} \cmidrule(lr){6-9} \cmidrule(lr){10-13} 
& \textbf{A} & \textbf{L} & \textbf{S} & \textbf{B} & \textbf{A} & \textbf{L} & \textbf{S} & \textbf{B} & \textbf{A} & \textbf{L} & \textbf{S} & \textbf{B} \\ 
\midrule 
GPT-5    & 11.11 & 25.00 & 20.78 & 21.33 &
68.52 & 74.19 & \underline{77.27 }&76.00 &
74.07 & 63.71 & 68.83 & 70.67 \\ 
o3                & 9.26 & 10.48 & 11.69 & 17.33 & \textbf{77.78} & \textbf{75.00} & \textbf{80.52} & \textbf{81.33} & \textbf{77.78} & \textbf{73.39} & \textbf{77.92} & \underline{78.67}\\ 
Claude-4.5 & 68.52 & \textbf{72.58} & \underline{72.73} & \textbf{78.67} & 72.22 & \underline{71.77} & 74.68 & \underline{78.67 }& 70.37 & \underline{72.58} & 74.68 & \underline{78.67} \\ 
Claude-3.7 & \underline{75.93} & 67.74 & 72.08 & 72.00 & \textbf{77.78} & 70.16 & 73.38 & 72.00 & 75.93 & \textbf{73.39} & 74.68 & 73.33 \\ 
DeepSeek-V3.2     & \textbf{79.63} & \underline{71.77} & \textbf{76.62} & \underline{77.33} &  \underline{74.07} & 70.97 & 74.03 & 76.00 & \underline{74.07} & 71.77 & \underline{77.27} & \textbf{80.00} \\ 
DeepSeek-R1       & \underline{75.93} & 69.35 & 71.43 & 68.00 & 72.22 & 72.58 & 72.73 & 77.33 & 75.93 & 67.74 & 70.13 & 73.33 \\ 
Gemini-2.5  & 24.07 & 32.26 & 31.82 & 30.67 & \textbf{77.78} & 69.35 & 72.08 & 77.33 & 70.37 & 67.74 & 71.43 & 72.00 \\ 
Qwen-3     & 11.11 & 10.48 & 11.04 & 12.00 & 18.52 & 24.19 & 18.18 & 16.00 & 27.78 & 22.58 & 20.78 & 18.67 \\ 
\midrule
\textbf{Average} & 44.45 & 44.96 & 46.02 & 47.17 & 67.36 & 66.03 & 67.86 & 69.33 & 68.29 & 64.11 & 66.97 & 68.17 \\
\bottomrule 
\end{tabular} 
\caption{Comparison of model performance across four problem complexity levels (A: Arithmetic, L: Logical, S: State, B: Boundary).
The process fidelity weight $W_s$ is reported.
The best performance is \textbf{bolded} and the second best is \underline{underline}.
\textbf{Overall, CoT and CoC substantially improve intermediate probe performance across all complexity dimensions compared to direct IO.}
} 
\label{tab:question_results} 
\end{table*}

\begin{figure}[!t]
    \centering
    \includegraphics[width=0.95\linewidth]{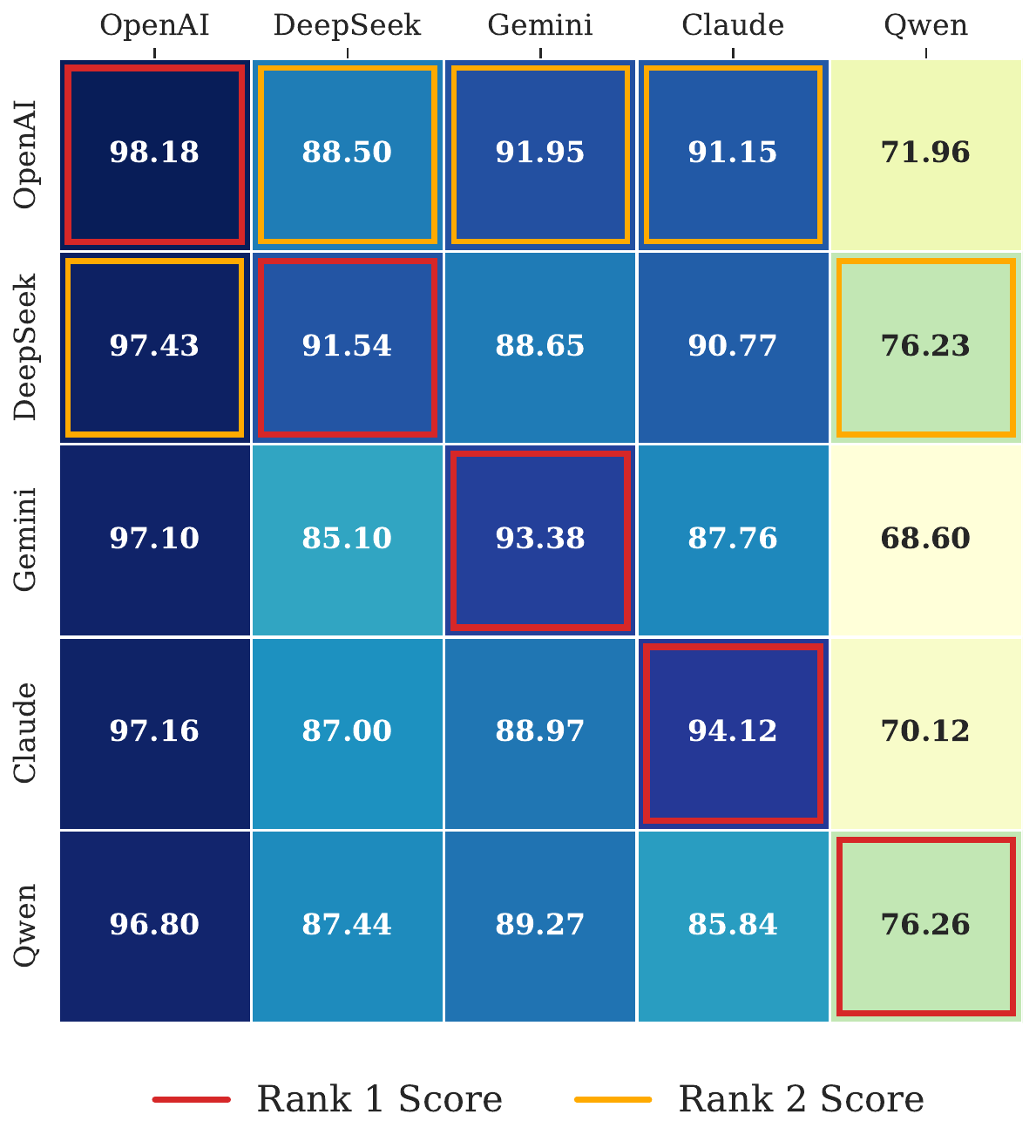}
    \caption{$I_s$ scores performance heatmap across five LLM families.
    Red and gold boxes indicate the rank-1 and rank-2 $I_s$ scores.
    \textbf{The pronounced diagonal pattern indicates a strong style bias.}
    }
    \label{fig:bias_heatmap}
\end{figure}

\section{Analysis}

To gain deeper insights into reasoning behavior in CoRE, we analyze several factors that influence their code reasoning performance.

\subsection{Robustness Gap and Style Overfitting}
\label{sec:robustness_gap}
To examine the detailed implementation invariance of LLMs in CoRE, we analyze their performance consistency across diverse LLM families.
Fig.\ref{fig:bias_heatmap} shows a heatmap of evaluation scores with a clear diagonal dominance, indicating that models typically exhibit the highest proficiency with code generated by their own model family.
For example, OpenAI models score 98.18 on OpenAI-generated implementations, noticeably higher than on code from other families.
This pattern suggests that current LLMs are strongly influenced by familiar implementation styles, indicating a tendency toward stylistic overfitting.
The lexical diversity of implementations, evidenced by a low 1-gram Jaccard similarity of $\mathcal{J}=0.6$, further highlights that surface-level differences can trigger this inconsistent reasoning.

\subsection{Investigating Superficial Execution}
A central contribution of the CoRE benchmark is the identification of superficial execution, defined as the phenomenon where models arrive at correct final outputs without accurately reasoning about intermediate states.
For instance, under the native IO prompting, GPT-5 achieves a high strict output accuracy $I=84.62$ but a remarkably low process fidelity weight $W_s$ of only 18.96.
This discrepancy yields a poor RCS score of $15.93$, as it penalizes such heuristic reasoning, proving that RCS effectively distinguishes genuine code reasoning from superficial execution.

\begin{figure}[!t]
    \centering
    \includegraphics[width=0.95\linewidth]{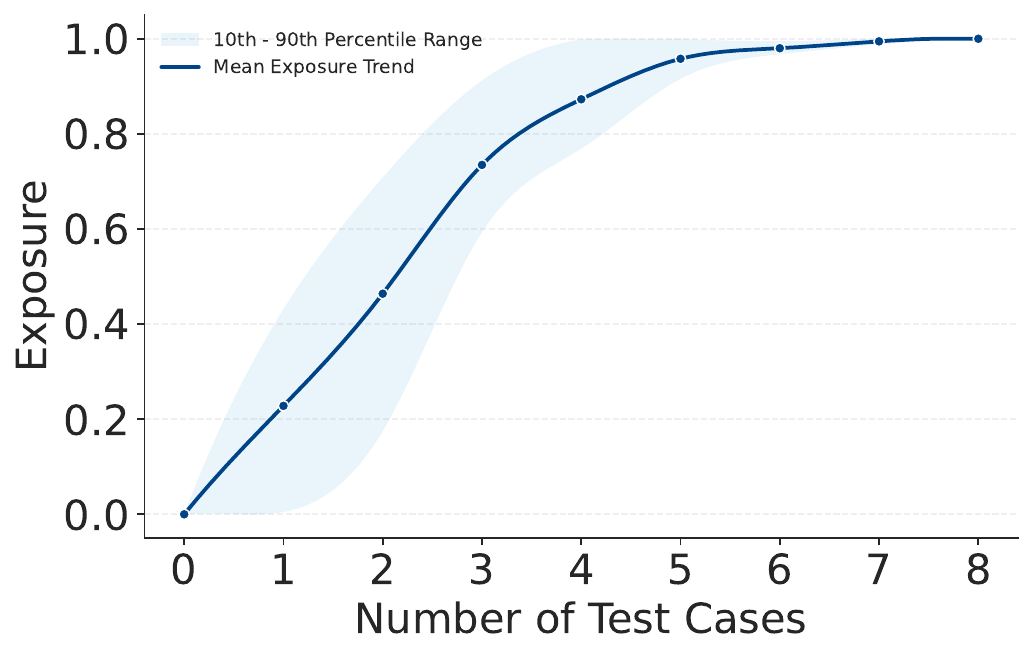}
    \caption{Exposure trend across increasing test cases. The solid blue line denotes the mean exposure ratio, while the shaded region represents the 10 $^{th}$ to 90$^{th}$ percentile distribution. \textbf{The model's performance shows initial variability before converging to saturation as the number of test cases increases.}}
    \label{fig:exposure}
\end{figure}

\subsection{Analysis of Dimensional Challenges}
To analyze how four dimensions challenge LLMs in process transparency evaluation, we report model performance across \textit{Arithmetic, Logic, State, and Boundary} in Tab.\ref{tab:question_results}.
The Arithmetic dimension exposes a severe lack of numerical precision.
Under IO prompting, GPT-5 and o3 achieve $W_S$ scores of only 11.11 and 9.26, respectively.
This suggests that native models often bypass exact arithmetic.
Structural fragility is evident in the limited performance of Logic and Boundary dimensions, which yield average $W_s$ of $44.96$ and $47.17$, respectively.
These results indicate that models often struggle to handle nested conditions and critical transition points.
Furthermore, the State dimension, which monitors variable histories across deep execution paths, shows an average $W_S$ of only $46.02$, revealing an inability to maintain a persistent internal memory of variable updates.
This poor performance stems from the fact that the State dimension frequently interacts with the other three dimensions as illustrated in Fig.\ref{fig:upset}, exposing their fundamental inability to maintain a persistent memory of variable updates.

\subsection{Impact of Prompting Method}
Although CoT, CoC, and RHDA enhance implementation invariance and process transparency evaluation, they still fail to address robustness gaps and superficial execution thoroughly.
Specifically, considering RHDA is an iterative reflection framework, we investigate why RHDA-2 sometimes underperforms RHDA-1 as observed in Tab.\ref{tab:main_results}.
We find that when a correct final output is mistakenly judged as incorrect in RHDA, models readily revise previously correct intermediate reasoning simply in response to this erroneous feedback, highlighting the existence of superficial execution.

\subsection{Soundness of Test Cases}
To improve computational efficiency, we employ a coverage-based reduction strategy to identify the minimal subset of test cases that maintains maximal execution coverage.
This optimization results in an average of 7.8 test cases per instance, which significantly raises the evaluation ceiling compared to prior benchmarks.
The exposure trend illustrated in Fig.\ref{fig:exposure} demonstrates that model performance initially varies but eventually converges to saturation as the number of test cases increases.

\section{Conclusion}
We introduce CoRE, a benchmark designed to evaluate code reasoning through implementation invariance and process transparency.
Our evaluation of eight leading LLMs identifies a pronounced robustness gap and the phenomenon of superficial execution in existing frontier LLMs.
To quantify this, we propose RCS, a composite protocol that penalizes heuristic shortcuts and rewards genuine code reasoning.
Our experimental findings highlight the fragility of current LLMs, underscoring the critical role of CoRE in facilitating a deeper exploration for code reasoning.

\section{Limitations}
Despite its rigor, the CoRE benchmark has several limitations that provide directions for future work.
First, our current implementation and execution behavior tracing are primarily focused on the Python programming language.
While CoRE derives instances from well-established benchmarks like HumanEval and LiveCodeBench, these source datasets may still be susceptible to potential training data leakage.
Second, although we incorporate a human verification stage to ensure logical validity, the reliance on expert verification limits the rapid and automated scaling of the dataset.
Finally, the distribution of probing questions across the four dimensions, Arithmetic, Logic, State, and Boundary, is not perfectly uniform due to the inherent complexity of intermediate execution.

\section{Acknowledgements}
This project is supported by the National Natural Science Foundation of China (No. 62302437) , and Yongjiang Talent Program (No. 2023A-402-G).

\bibliography{custom}

\appendix
\clearpage
\begin{table}[h]
\centering
\caption{Overview of Large Language Models used in different stages of the study. \textbf{II} indicates implementation invariance, and \textbf{IP} represents intermediate probing.}
\label{tab:model_selection}
\begin{tabular}{lccc}
\toprule
\textbf{Model} & \textbf{II} & \textbf{IP} & \textbf{Eval} \\
\midrule
GPT-5            &            & \checkmark & \checkmark \\
o3           &            &            & \checkmark \\
o1           & \checkmark &            &            \\
4o-mini      & \checkmark &            &            \\
DeepSeek-V3      & \checkmark &            &            \\
DeepSeek-V3.2    &            & \checkmark & \checkmark \\
DeepSeek-R1      & \checkmark &            & \checkmark \\
Claude-3.7       & \checkmark &            & \checkmark \\
Claude-4.5       &            & \checkmark & \checkmark \\
Qwen-3-235B       & \checkmark & \checkmark & \checkmark \\
Gemini-2.5       & \checkmark & \checkmark & \checkmark \\

\bottomrule
\end{tabular}

\end{table}

\begin{figure}
    \centering
    \includegraphics[width=0.95\linewidth]{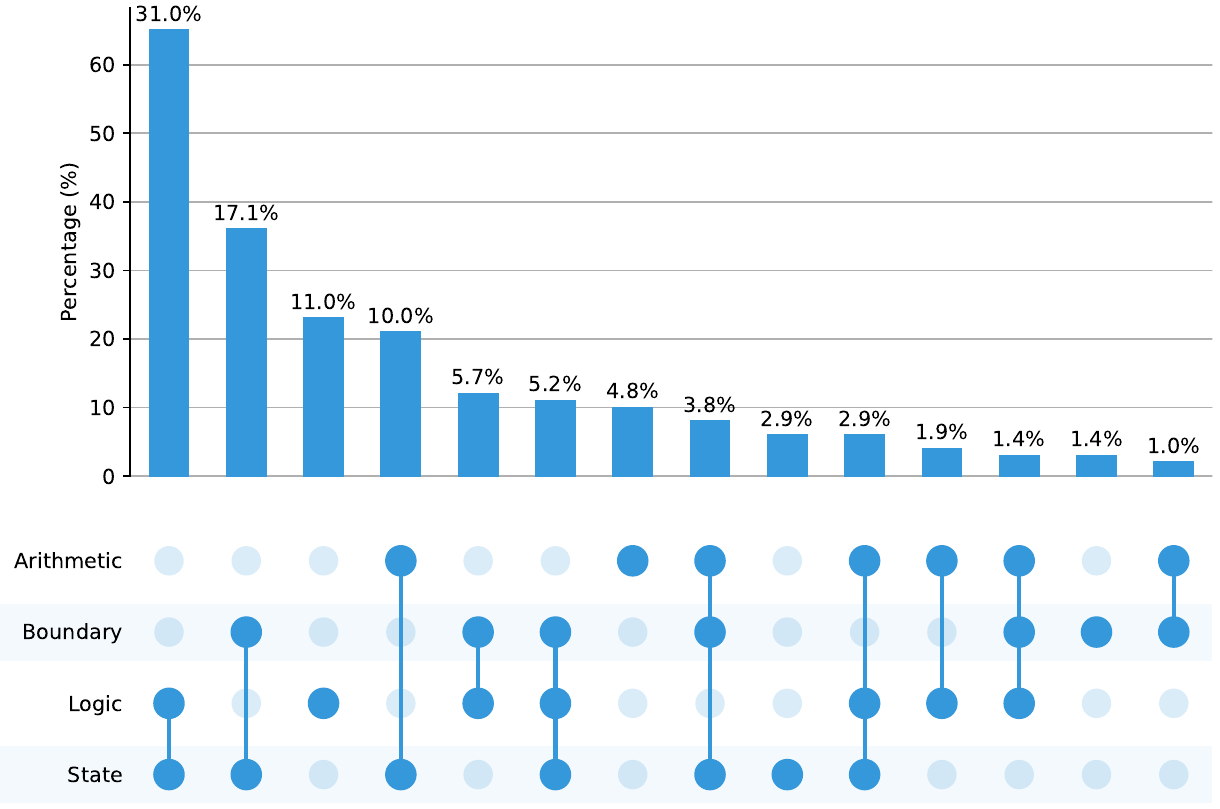}
    \caption{The interaction of intermediate probes.}
    \label{fig:upset}
\end{figure}

\section{Benchmark Construction Details}
\subsection{Details of Identical Implementation Generation}\label{app:implement}
We leverage an ensemble of state-of-the-art LLMs to generate diverse but functionally equivalent code implementations for problems sourced from HumanEval and LiveCodeBench.
As shown in Table \ref{tab:model_selection}, seven models, such as o1, 4o-mini, DeepSeek-V3, and Qwen-3, were primarily utilized for this stage.
During this stage, we set the temperature to 0.7.
In the implementation cleaning process, the cyclomatic complexity and Jaccard similarity threshold $\tau_{cc}$ and $\tau_{sim}$ are set to 3.0 and 0.7, respectively.

\begin{promptbox}{Code Generation Prompt}
You are an expert Python programmer. You will be given a question (problem specification) and will generate a correct Python program that matches the specification and passes all tests.\\[8pt]

\textbf{\#\#\# Question:} \\
\{question\_content\} \\[8pt]

\textbf{\#\#\# Format:} You will use the following starter code to write the solution to the problem and enclose your code within delimiters. \\
\texttt{```python} \\
\{starter\_code\} \\
\texttt{```} \\[8pt]
\textbf{\#\#\# Answer: (use the provided format with backticks)}
\end{promptbox}

\subsection{Details of Intermediate Probing}\label{app:probing}
To alleviate the impact of LLM bias, we employ GPT-5, DeepSeek-V3.2, Claude-4.5, DeepSeek-R1, and Qwen-3 in a cycle to formulate probing questions in this stage.
We encourage questions to be diverse to challenge LLMs, and we thus slightly increase the temperature to 0.8.
The following few-shot examples are provided to the model to demonstrate the desired reasoning depth and output format.
\begin{promptbox}{Reference Examples for Code Reasoning}
\label{app:few_shot_examples}
\textbf{\#\# Example 1} \\
\textbf{Input Source Code:}
\begin{lstlisting}[language=Python]
def update_arr(arr, factor):
  for i in range(len(arr)):
    if i > 0:
      arr[i] = (arr[i] * 
      factor + arr[i-1]) % 10
  return arr
\end{lstlisting}

\textbf{Input Execution Trace:}

[Step 1] Line 2: for i in range(len(arr)) (Iter 1: i=0) | locals: {'arr': [1, 5, 9], 'factor': 2}

...

[Step 5] Line 3: if i > 0 (Evaluated: True)

[Step 6] Line 5: arr[i] = (arr[i] * factor + arr[i-1]) \% 10 | locals: {'arr': [1, 1, 9]}

\textbf{Output (JSON):}
\begin{lstlisting}[language=json]
{
  "target_step_index": 6,
  "complexity_type": "Arithmetic + State",
  "question": "In the 2nd iteration (i=1), what is the exact value assigned to 'arr[1]' on line 5?",
  "ground_truth": "1",
}
\end{lstlisting}

\textbf{\#\# Example 2} \\
\textbf{Input Source Code:}
\begin{lstlisting}[language=Python]
def matrix_sum(matrix):
  total = 0
  for row in matrix:
    for val in row:
      if val == -1: break
      total += val
    return total
\end{lstlisting}

\textbf{Input Execution Trace: }

...

[Step 10] Line 3: for row in matrix (Iter 2) | locals: \{'row': [10, -1, 5], 'total': 6\} 

[Step 11] Line 4: for val in row (Iter 1) | locals: \{'val': 10\}

[Step 12] Line 5: if val == -1 (Evaluated: False)

[Step 13] Line 6: total += val | locals: \{'total': 16\}

[Step 14] Line 4: for val in row (Iter 2) | locals: \{'val': -1\}

[Step 15] Line 5: if val == -1 (Evaluated: True)

[Step 16] Line 5: break | locals: \{'total': 16\}

\textbf{Output (JSON):}
\begin{lstlisting}[language=json]
{
  "target_step_index": 16,
  "complexity_type": "Boundary + State",
  "question": "Given the input row [10, -1, 5], what is the value of variable 'total' immediately after the break statement executes on line 5?",
  "ground_truth": "16",
}
\end{lstlisting}
\end{promptbox}
Then we combine the above reference examples with the following instruction, providing LLMs with sample questions targeting four dimensions: Arithmetic, Logic, State, and Boundary
\begin{promptbox}{Intermediate Probing Prompt}
\textbf{\# System Instruction} \\
You are an expert Code Logic Auditor. Your goal is to create a ``Challenging Code Reasoning Benchmark'' by analyzing execution traces.

\medskip
\textbf{\#\# Mission:} \\
Identify the \textbf{most error-prone} step in the provided execution trace and formulate a question about it. We want to test the model's ability to handle \textbf{complexity}.

\medskip
\textbf{\#\# Selection Criteria:}
\begin{enumerate}
    \item \textbf{Complex Arithmetic:} Lines with multiple operators (e.g., \texttt{res = (a + b) * c \% d}) or list indexing with math.
    \item \textbf{Loop Boundaries:} The last iteration of a loop, or the state immediately after a loop finishes.
    \item \textbf{Nested Logic:} Steps inside a nested loop or a nested \texttt{if} block where context is deep.
    \item \textbf{Compound Conditions:} Boolean evaluations involving \texttt{and, or, not}.
    \item \textbf{State Accumulation:} A variable modified multiple times (e.g., \texttt{total} after the 5th iteration).
\end{enumerate}

\medskip
\textbf{\#\# Rules:}
\begin{itemize}
    \item \textbf{Answer Integrity:} The `ground\_truth` must be extracted EXACTLY from the provided trace. Do not compute it yourself.
    \item \textbf{Precision:} The question must specify the exact context (e.g., "At the end of the 3rd iteration...", "In the evaluation of the condition on line 5...").
\end{itemize}

\textbf{\# Few-Shot Examples} \\
\textit{\textlangle Reference Examples from Appendix~\ref{app:few_shot_examples} are inserted here\textrangle}  \\

\textbf{\# Current Task}

\textbf{Input Source Code:} \\
\texttt{```python} \\
\{source\_code\} \\
\texttt{```} \\[8pt]
\textbf{Input Execution Trace:}\\
\{execution\_trace\}
\end{promptbox}

\subsection{Details of Human Verification}\label{app:human}
To ensure the logical validity and diversity of the benchmark, five experts executed a multi-stage verification protocol.
During the implementation variance phase, the experts manually refined samples with a Jaccard similarity within the range of $[0.7, 0.9]$ to recover implementations that exhibited high 1-gram similarity but featured distinct structural logic or algorithmic strategies.

Regarding intermediate reasoning traces, the experts evaluated logical correctness and the rationality of state transitions while simultaneously removing redundant questions to maximize diversity.
This rigorous calibration process required approximately 110 human-hours and resulted in a finalized pool of 255 expert-verified implementations and 243 intermediate questions across 60 unique questions.
The effort achieved high inter-annotator agreement with Cohen's Kappa $\kappa=0.82$ for the implementation variance phase and $\kappa=0.86$ for the intermediate probing phase.

\section{Details of Experimental Baselines}\label{app:baseline}

\subsection{Prompting Methods}
We evaluated the models using three primary strategies: Input-Output (IO), Chain-of-Thought (CoT), and Chain-of-Code (CoC). For CoT, we specifically implemented a two-shot approach to standardize the reasoning "scratchpad" across all frontier models.

\begin{promptbox}{Standard Output Prompt}
You are given a Python function and an assertion containing an input to the function.
Complete the assertion with a literal (no unsimplified expressions, no function calls) containing the output when executing the provided code on the given input, even if the function is incorrect or incomplete.
Provide the full assertion with the correct output, following the examples.

\begin{lstlisting}[language=Python]
def f(s):
  return s + "a" 
assert f("x9j") == ??
# Answer:
assert f("x9j") == "x9ja"

{code}
assert {func_name}({x}) == ??
# Answer:
\end{lstlisting}
\end{promptbox}

\begin{promptbox}{Chain-of-Thought Prompt}
You are given a Python function and an assertion containing an input to the function. Complete 
the assertion with a literal (no unsimplified expressions, no function calls) containing the output when executing 
the provided code on the given input, even if the function is incorrect or incomplete. Execute the program step by 
step before arriving at an answer, and provide the full assertion with the correct output, following the examples.

\begin{lstlisting}[language=Python]
def f(s):
    s = s + s
    return "b" + s + "a"
assert f("hi") == ??
\end{lstlisting}
Let's execute the code step by step:
\begin{itemize}
\item 1. The function f is defined, which takes a single argument s.
\item 2. The function is called with the argument "hi", so within the function, s is initially "hi".
\item 3. Inside the function, s is concatenated with itself, so s becomes "hihi".
\item 4. The function then returns a new string that starts with "b", followed by the value of s (which is now "hihi"), and ends with "a".
\item 5. The return value of the function is therefore "bhihia".
\end{itemize}

\begin{lstlisting}[language=Python]
Answer:
assert f("hi") == "bhihia"
\end{lstlisting}
\medskip
\begin{lstlisting}[language=Python]
{code}
assert {func_name}({x}) == ??
\end{lstlisting}
Let's execute the code step by step:
\end{promptbox}

\begin{promptbox}{Chain-of-Code Prompt}
You are given a Python function and an assertion containing an input to the function. Complete 
the assertion with a literal (no unsimplified expressions, no function calls) containing the output when executing 
the provided code on the given input, even if the function is incorrect or incomplete. Execute the program step by 
step before arriving at an answer, and provide the full assertion with the correct output, following the examples.

\begin{lstlisting}[language=Python]
def f(s):
    s = s + s
    result = "b" + s + "a"
    return result
assert f("hi") == ??
\end{lstlisting}
\texttt{[TRACE]}\\
state: \{\} \\
line: f("hi") \\
explanation: Python execution. \\
delta state: {{'s': 'hi'}} \\
line: s = s + s \\
explanation: Python execution. \\
delta state: {{'s': 'hihi'}} \\
line: result = "b" + s + "a" \\
explanation: Python execution. \\
delta state: {{'result': 'bhihia'}} \\
line: return result \\
explanation: Python execution. \\
delta state: {{}} \\
\texttt{[/TRACE]} \\
Answer:
\begin{lstlisting}[language=Python]
assert f("hi") == "bhihia"
\end{lstlisting}

\begin{lstlisting}[language=Python]
{code}
assert {func_name}({x}) == ??
\end{lstlisting}
\texttt{[TRACE]}
\end{promptbox}
\end{document}